\documentclass{emulateapj}

\slugcomment{Accepted for publication in ApJ}

\newcommand{\cs}{c_{\rm s}}
\newcommand{\degK}{\,{}^\circ\mbox{K}}

\newcommand{\eff}{\mathrm{eff}}

\newcommand{\kB}{k_{\rm B}}
\newcommand{\kms}{\mbox{\,km s}^{-1}}
\newcommand{\LJ}{L_{\rm J}}
\newcommand{\Mbox}{M_{\rm box}}

\newcommand{\Myr}{\mbox{\,Myr}}
\newcommand{\pcc}{\mbox{\,cm}^{-3}}
\newcommand{\pc}{\mbox{\,pc}}
\newcommand{\rhoc}{\rho_{\rm c}}
\newcommand{\rhoj}{\rho_{\rm J}}

\newcommand{\MJeff}{M_{{\rm J},\eff}}

\newcommand{\dif}{\mathrm{d}}

\newcommand{\BP}{Ballesteros-Paredes}
\newcommand{\VS}{V\'azquez-Semadeni}
\newcommand{\Pext}{P_{\rm ext}}
\newcommand{\tff}{\tau_{\rm ff}}

\newcommand{\Eg}{E_{\rm g}}
\newcommand{\Eth}{E_{\rm th}}
\newcommand{\Jsq}{J^2}
\newcommand{\gamef}{\gamma_{\rm eff}}
\newcommand{\gamefc}{\gamma_{\rm eff,c}}
\begin{document}

\title{Formation and Collapse of Quiescent Cloud Cores Induced by
Dynamic Compressions}

\author{Gilberto C. G\'omez\altaffilmark{1}, Enrique
V\'azquez-Semadeni\altaffilmark{1}, Mohsen Shadmehri\altaffilmark{2,3} and
Javier Ballesteros-Paredes\altaffilmark{1}}

\altaffiltext{1}{Centro de Radioastronom\1a y Astrof\1sica, Universidad
Nacional Aut\'onoma de M\'exico,
       Apartado Postal 3-72 (Xangari),
       Morelia, Mich. 58089, M\'exico}
\email{g.gomez, e.vazquez, j.ballesteros@astrosmo.unam.mx}

\altaffiltext{2}{Department of Physics, School of Science, Ferdowsi
University, Mashhad, Iran}

\altaffiltext{3}{School of Mathematical Sciences, Dublin City
University, Glasnevin, Dublin 9, Ireland} 
\email{mohsen.shadmehri@dcu.ie}

\begin{abstract}

We present numerical hydrodynamical simulations of the formation, evolution and
gravitational collapse of isothermal molecular cloud cores induced by
turbulent compressions in spherical geometry. 
A compressive wave is set up in a constant sub-Jeans
density distribution of radius $r =1$ pc.
As the wave travels through the simulation grid, a shock-bounded
spherical shell is formed.
The inner shock of this shell reaches and bounces off the center,
leaving behind a central core with an initially almost uniform density
distribution, surrounded by an envelope consisting of the material in
the shock-bounded shell, with a power-law density profile
that at late times approaches a logarithmic slope
of $-2$ even in non-collapsing cases. The central core and the
envelope are separated 
by a mild shock. The resulting density structure resembles a quiescent
core of radius $\lesssim 0.1$ pc, with a
Bonnor-Ebert-like (BE-like) profile, although it has significant dynamical
differences: it is initially non-self-gravitating and confined by the
ram pressure of the infalling material, and consequently, 
growing continuously in mass and size. 
%Since the gas in the layer keeps falling inward, flowing through the
%inner shock, the core mass increases.
With the appropriate parameters, the core mass eventually reaches an
effective Jeans mass, at which time the core begins to 
collapse, until finally a
singularity is formed. Thus, there is necessarily a
time delay between the appearance of the core and the onset of its
collapse, but this is not due to the dissipation of its internal
turbulence as it is often believed, but rather to the time necessary for
it to reach its Jeans mass. These results
%generates 
suggest that
pre-stellar cores may approximate Bonnor-Ebert 
structures which are however
of variable mass and may or may
not experience gravitational 
collapse, in qualitative
agreement with the large observed frequency of
cores with BE-like profiles.
In our collapsing simulations, a time $\sim 0.5$ Myr
typically elapses between the
formation of the core and the time at which it becomes gravitationally
unstable, and another $\sim 0.5$ Myr are necessary for it to complete
the collapse.

\end{abstract}


%-----------------------------------------------------------------

\section{INTRODUCCION}
\label{sec:intro}

The process by which a gas parcel (``core'') within a molecular cloud (MC)
initiates a collapse leading to the formation of a star or group of
stars remains loosely understood, in particular
the details of its dynamical
evolution. Observations indicate that ``prestellar'' molecular cloud
cores (i.e., those that do not yet contain a protostellar object, but
that appear to be on route to forming it) have a density structure that
resembles Bonnor-Ebert (BE) profiles \citep{Ebert55, Bonnor56},
being nearly flat in their central regions, while approaching
the singular isothermal sphere (SIS) profile $n(r) \propto
r^{-2}$ at large radii. ``Stellar'' cores (those already containing a
Class 0 or Class 
I protostellar object), on the other hand, appear to have density
profiles closer to that of the SIS throughout their volume
\citep[e.g.,][]{ALL01,Cas02,KirkJ05,Lee_etal07} \citep[see also the
reviews by][ and references therein]{LAL07,diFran07,WardT07}. 

The line profiles and spatial distribution of molecular-line
observations provide further clues to the dynamics. For example,
based on observations of CS(3-2), CS(2-1), DCO$^+$(2-1) and
N$_2$H$^+$(1-0), \citet{LMP04} found a moderate fraction of prestellar
cores (18 out of 70 in their Table 2) showing clear evidence of
subsonic inward radial motions, at velocities $v \lesssim 0.07 \kms$.
Moreover, studies of individual starless cores have suggested that
the radial velocity does not increase appreciably towards the center
\citep{Tafalla_etal98, WMWdF99, TMCW04, Lee_etal07, Schnee_etal07}.  Those inward
motions frequently extend to long enough distances from the cores'
centers (a few tenths of a parsec) that they seem inconsistent with
the ``inside-out'' collapse model of \citet{Shu77}, since a central
protostar should have had time to form by the time the rarefaction
wave reaches those distances \citep[][ and references
therein]{diFran07}.

A large number of theoretical studies have investigated the collapse
process starting from a variety of initial and boundary conditions, both
analytically, through similarity solutions, and numerically
\citep[e.g.,][]{Larson69, Penston69a, Penston69b, Shu77, Hunter77, FC93,
HWGA03}. All of these studies have considered the collapse of a
\emph{fixed} mass of gas, either through the usage of a hot, tenuous
confining medium that pressure-confines the core while adding no weight
to it, or through fixed boundaries. Moreover, most of these studies used
static initial configurations, either with uniform density or with
BE hydrostatic equilibrium profiles.

On the other hand, MCs are thought to be supersonically turbulent,
since they
exhibit supersonic linewidths \citep{ZP74}, and MC cores,
as well as their parent MCs themselves, have been suggested to be
turbulent density fluctuations within their environments \citep{Sasao73,
Elm93, BVS99},
being produced by effectively supersonic compressions. \citet{HF82}
showed that the effective Jeans mass of a fluid parcel subject to an
external compressive velocity field is significantly decreased with
respect to its normal static value. Furthermore,
\citet{VKSB05} have recently pointed out that, if MCs are
isothermal throughout,\footnote{
  %Note that recent numerical evidence suggests that
  %MCs may not be completely occupied by isothermal molecular gas, but
  %rather consist of molecular clumps embedded in a warmer, atomic
  %interclud medium \citep[e.g.,][]{HI06,VGJBG07}.
  The possibility that MCs may contain warmer atomic gas interspersed
  with the colder molecular gas has been recently raised by
  \citet{HI06}.
  If this turns out to be
  the case, then the argument against the feasibility of stable
  hydrostatic BE-like MC cores weakens significantly.}
then the hot,
tenuous medium necessary to confine and stabilize a hydrostatic
equilibrium configuration is not available, and the equilibrium state is
then expected to be unstable in general.

This leads naturally to the question of whether hydrostatic equilibrium
configurations can be produced in such turbulent conditions, and if so,
how do they arrive at that state. Otherwise, if the entire process is
dynamic, one can ask what is the density and velocity structure of the
cores at the time they engage into collapse and, if different from the
initial conditions normally assumed, what effects does it have on the
evolution. Moreover, if the
%collapse is triggered
core is formed and induced to collapse by a compressive wave,
then in general there is an \emph{inflow} that builds up the core
dynamically, and the mass that ends up collapsing is not previously
determined by the initial conditions, but rather is determined ``on the
spot'' depending on the local instantaneous conditions. The studies of
collapse mentioned above cannot answer these questions, since they
already assume gravitationally unstable structures, and initial
hydrostatic equilibrium configurations, so that all of the mass is
involved in the collapse. 

A study that comes close to these
goals is that by \citet{HWGA03}, 
who investigated the effect of increasing in the pressure external
$\Pext$ to an initially stable BE sphere. They noted that the
resulting configurations are a good match to the observations because
the density profile is flat at the center, and the prestellar phase is
characterized by subsonic inwards velocities at the outskirts, and
by
nearly zero velocity at the inner parts. However, having a hot confining
medium outside and an initial hydrostatic profile, this study still
could not capture the core \emph{formation} part of the evolution, and 
predetermined the mass that collapses from the initial conditions. Also, it
did not consider the possibility of a transient compression and thus of
a failure to collapse.

A brief discussion of the dynamic scenario of core
formation has been given by \citet{Whit_etal07} in the context of the
formation of cores that give birth to brown dwarfs. These authors have
suggested that the dynamic formation of cores should involve a mass growth
period and the confinement of BE-like structures by ram pressure of
external infalling material. 

In view of the above, in this paper we then present
numerical hydrodynamical simulations 
in spherical coordinates of transient compressions in 
homogeneous, initially gravitationally stable regions,
with the purpose of investigating the formation of cores embedded in
turbulent molecular clouds. In particular, we focus on the evolution
of its density and velocity profiles, the timescales required for a core
to be assembled and then collapse or redisperse, and the mechanism by
which a certain fraction of the mass is gravitationally ``captured''
to then proceed to collapse.

In particular, the timescale issue is highly relevant because it is
often
thought that the prestellar lifetimes of the cores in the turbulent
scenario of star formation are of the order of \emph{one}
core's free-fall time $\tff$.
However, it has been shown by \citet{VKSB05} and \citet{GVKB07}
that even in highly dynamical, driven-turbulence simulations, the lifetimes
are a few to several times $\tff$. It is important then to investigate
the detailed evolution of cores formed by turbulent compressions, to
understand the reason for those observed timescales.

The plan of the paper is as follows. In \S \ref{sec:model}
we first discuss the motivation and applicability of the
spherical symmetry used in this paper, and then we
describe
the numerical setup of the problem.
In \S \ref{sec:results}
we present the results of two fiducial cases of core evolution, one
collapsing and one rebounding, and in \S \ref{sec:discussion} we then
disccus the implications of our results, and compare with existing
observational and theoretical work. 
Finally, in \S \ref{sec:conclusions}
we present a summary and some concluding remarks.

%-----------------------------------------------------------------

\section{THE MODEL}
\label{sec:model}

\subsection{The need for focused compressions} \label{sec:focusing}

The formation, and subsequent induction to
gravitational collapse, of clouds and clumps by compressive
velocity fields  
(as opposed, in particular, to collisions of
pre-existing clouds) has 
been studied intensively by numerous workers for more than three decades
\citep[e.g.,][]{Sasao73, HF82, TBC87, MZGH93, Elm93, VPP96, VBK03,
VKSB05, VGJBG07, BVS99, OGS99, OSG01, KHM00, HMK01, LNMH04, TP04,
TP05}.
In such scenario, the formation of cores and stars is in
agreement with observational studies that suggest that star formation
is a rapid and dynamic process \citep[e.g.,][]{Lee_Myers99, BHV99,
  Elmegreen00, Pringle_etal01, Briceno_etal01, HBB01, BH07}.
The ability of a compression to induce
collapse is directly related to the stability of 
self-gravitating equilibrium structures, which in turn depends
critically on the geometry of the configurations and on the {\it
effective polytropic exponent} ($\gamef$) of the medium.
This exponent
describes the response to compressions of a medium subject to heating
and cooling processes \citep{TBC87, Elm91, VPP96}, so that the flow exhibits an
effective polytropic equation of state of the form $P \propto
\rho^{\gamef}$. 

The stability of self-gravitating structures depends both on the
geometry and on $\gamef$ because both of them influence the variation of
the ratio $\Jsq \equiv |\Eg|/\Eth$ upon compressions, where $|\Eg|$ is the
absolute value of the gravitational energy and $\Eth$ is the
(supporting) internal energy. For example, it is well known that the
existence of stable {\it spherical} configurations without any external
confining agent requires $\gamef > 4/3$ \citep[][ \S 117]{Chandra61}.
In this case, $\Jsq$ decreases upon a compression, and increases upon an
expansion, rendering the equilibrium stable. This behavior is reversed
for $\gamef < 4/3$, so that equilibrium configurations are unstable in
this case. We refer to the value of $\gamef$ at which the reversal
occurs as the {\it critical} value, $\gamefc$.

Now, if the compression occurs along $\nu$ directions, so that the
density increases as $L^\nu$, where $L$ is the length scale along the
direction(s) of compression, then the rate of variation of $|\Eg|$ and
of $\Jsq$ with the compression depends on $\nu$ and, as a consequence,
the critical value of $\gamef$ also depends on $\nu$. Specifically, one
obtains \citep{MZGH93, VPP96}
\begin{equation}
\gamefc = 2(1 - \nu^{-1}).
\label{eq:gamefc}
\end{equation}
This recovers the value $\gamefc = 4/3$ for
three-dimensional (e.g., spherically symmetric) compressions, as well as
the well known fact that planar compressions ($\nu =1$) cannot induce
collapse in isothermal flows ($\gamef =1$), since $\gamefc = 0$ in this
case. Equation (\ref{eq:gamefc}) also shows that the induction of
collapse in isothermal media requires compressions in more than two
dimensions ($\nu > 2$).

These considerations show that
useful insight can be gained from the analysis of
spherically symmetric compressions, like those
assumed in this paper, since the
compressions that induce the collapse of selected subregions of
turbulent, isothermal molecular clouds need to be of dimensionality
higher than 2. Such compressions are expected to be rare but still
existing in general supersonic turbulent regimes, involving a certain
degree of \emph{focusing} (or \emph{convergence}) of the flow.
For example,
\citet{Whit_etal07} appeal to the low but finite probability of such
focused compressions to explain the scarcity of $\sim 0.01 M_\odot$
brown dwarfs. 

Finally, note that these convergent flows are in principle
different from a simple passing shock (which is essentially a planar
compression), although the latter can also induce multidimensional
compression when the shock has a finite transverse extent, producing a
flattened structure that then can contract gravitationally in the
tranverse direction.

%-----------------------------------------------------------------

\subsection{The numerical setup}
\label{sec:num_setup}

In view of the above considerations, in
this work we consider an idealized spherical
cloud subject to an external
compression wave. The simple spherical
geometry allows us to focus on the basic phenomena related to the
effects of the compression. Moreover,
by considering a uniform density distribution and allowing the
system to dynamically choose the amount of mass involved in the
collapse, we  avoid some of the restrictions that previous
work has imposed on the evolution.

The hydrodynamic evolution of this setup
was solved using ZEUS \citep{sto92},
a finite difference, time explicit, operator split, hydrodynamic code.
The calculations were performed on a 1D spherical grid,
with the domain spanning the range $0 < r < 1 \pc$
with 1000 grid points 
spaced such that $\delta r_{i+1} / \delta r_i = 1.005$.
This yields a spatial resolution of $\approx 3 \times 10^{-5} \pc$
at the inner boundary, and $\approx 5 \times 10^{-3} \pc$ at the
outer boundary.
(Selected simulations were 
performed with a much higher resolution of 4000 grid points and no
significant differences were observed.)
The boundary conditions are ``reflecting'' at $r=0$ and ``outflow''
at $r=1 \pc$.
\emph{No confining agents are used whatsoever}
(closed boundary nor hot tenuous medium),
implying that mass can freely leave the system, although
it cannot enter. The absence of a confining agent attempts to emulate
the situation of a density enhancement immersed in a much more extended
medium at the same temperature.

All simulations started with a constant density distribution,
an isothermal equation of state, and
a temperature $T = 11.4 \degK$ which, with a mean particle mass
$\mu = 2.36~m_{\rm H}$, yields an isothermal sound speed $c_s = 0.2 \kms$.
This setup was perturbed by a compressive velocity pulse
given by the relation

\begin{equation}
  v(r) = \cases{
           \quad 0, & $r < r_0-dr_0$; \cr
          -v_0 \, \sin\left(\frac{\pi}{2} \frac{r-r_0}{dr_0}\right),
             & $r_0 - dr_0 < r < r_0 + dr_0$; \cr
          -v_0 \, \sin\left(\frac{\pi}{2} \frac{r_1-r}{dr_1}\right),
             & $r_0 + dr_0 < r$; \cr
         }
\label{eq:pulse}
\end{equation}

\noindent
where $v_0$ and $r_0$ are parameters of the simulation,
$dr_0 = 0.1 \pc$, $r_1 = [ r_{\rm max} + ( r_0 + dr_0 ) ] / 2$, 
and $dr_1 = [ r_{\rm max} - ( r_0 + dr_0 ) ] / 2$,
with $r_{\rm max} = 1\pc$.
A simple self-gravity module was also added to the code.

Our approach continues along the lines of simple, basic models
that have explored the gravitational collapse of MC cores,
since \citet{Larson69} and \citet{Penston69a,Penston69b},
through \citet{HWGA03}, which we have extended to include
an initial velocity impulse,  intended to mimic
the random compressive motions expected in a turbulent medium.
Nevertheless, the one-dimensional nature of the model, together with
the adopted spherical geometry, makes this setup somewhat unphysical
as it restricts the nature 
of the compressible wave to spherical shells, and 
``turbulent'' support in this model is present only as purely divergent
motion, with no rotational component.
A more realistic way of modeling the core formation process,
albeit perhaps less amenable to detailed analysis, would be to 
perform full 3D numerical simulations, via random compressions of finite
cross-section 
generated by bulk motions of the gas, similarly to what has been done
for the diffuse medium by  \citet{VGJBG07}. We intend to pursue this in
the near future, over the theoretical foundation laid out by the
simple present study.

Another limitation introduced by the adopted geometry is the large
mass of the collapsed core resulting from our simulations
(cf. \S \ref{sec:induced_col}).
In a more realistic simulation, without the geometrical and symmetry
restrictions, the collapsing system would probably undergo
fragmentation.
Therefore, we see the collapsed objects generated in these
simulations not as a single star,
but as the precursors of small clusters.

%-----------------------------------------------------------------

\begin{figure*}[!t]
\plottwo{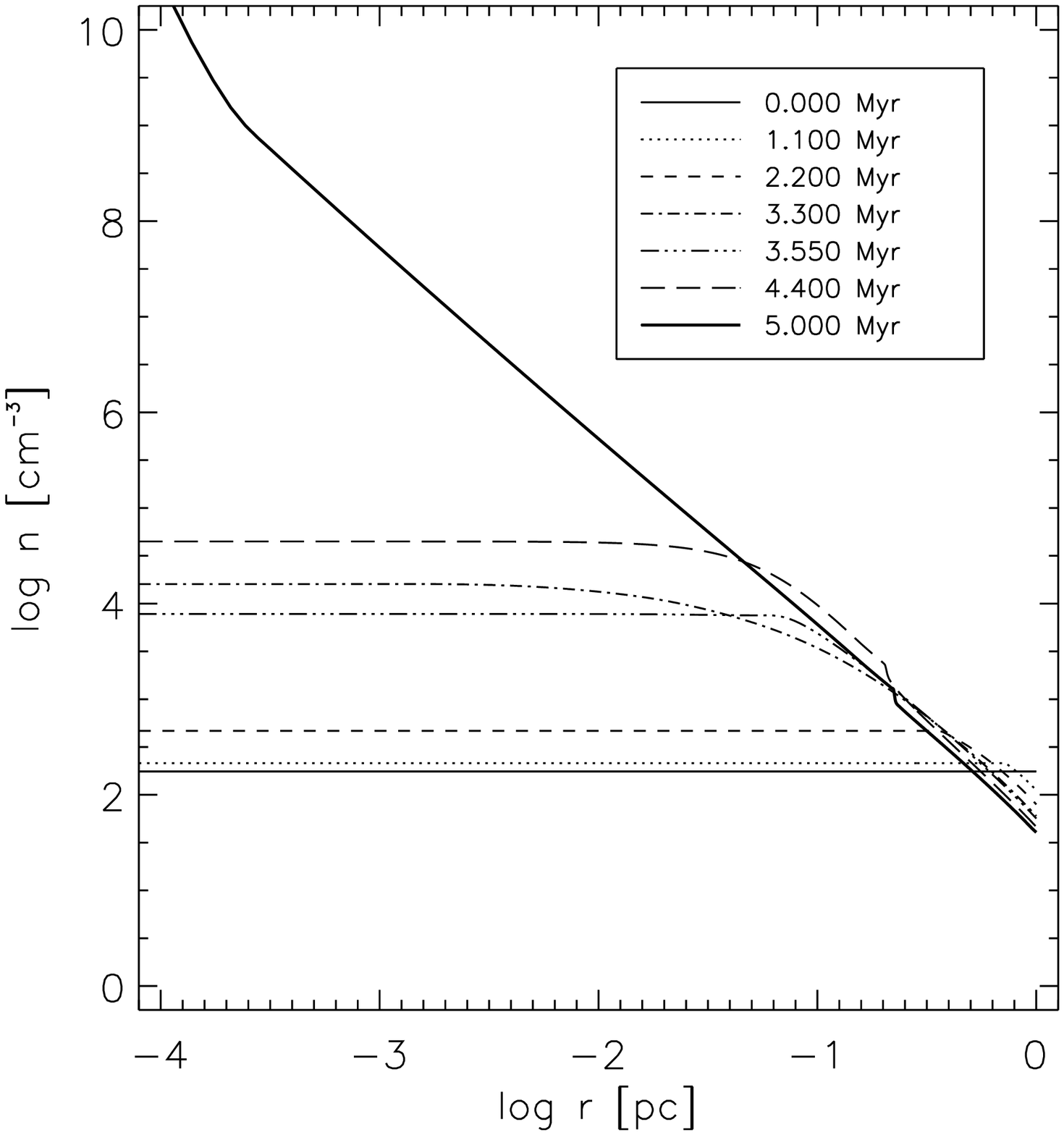}{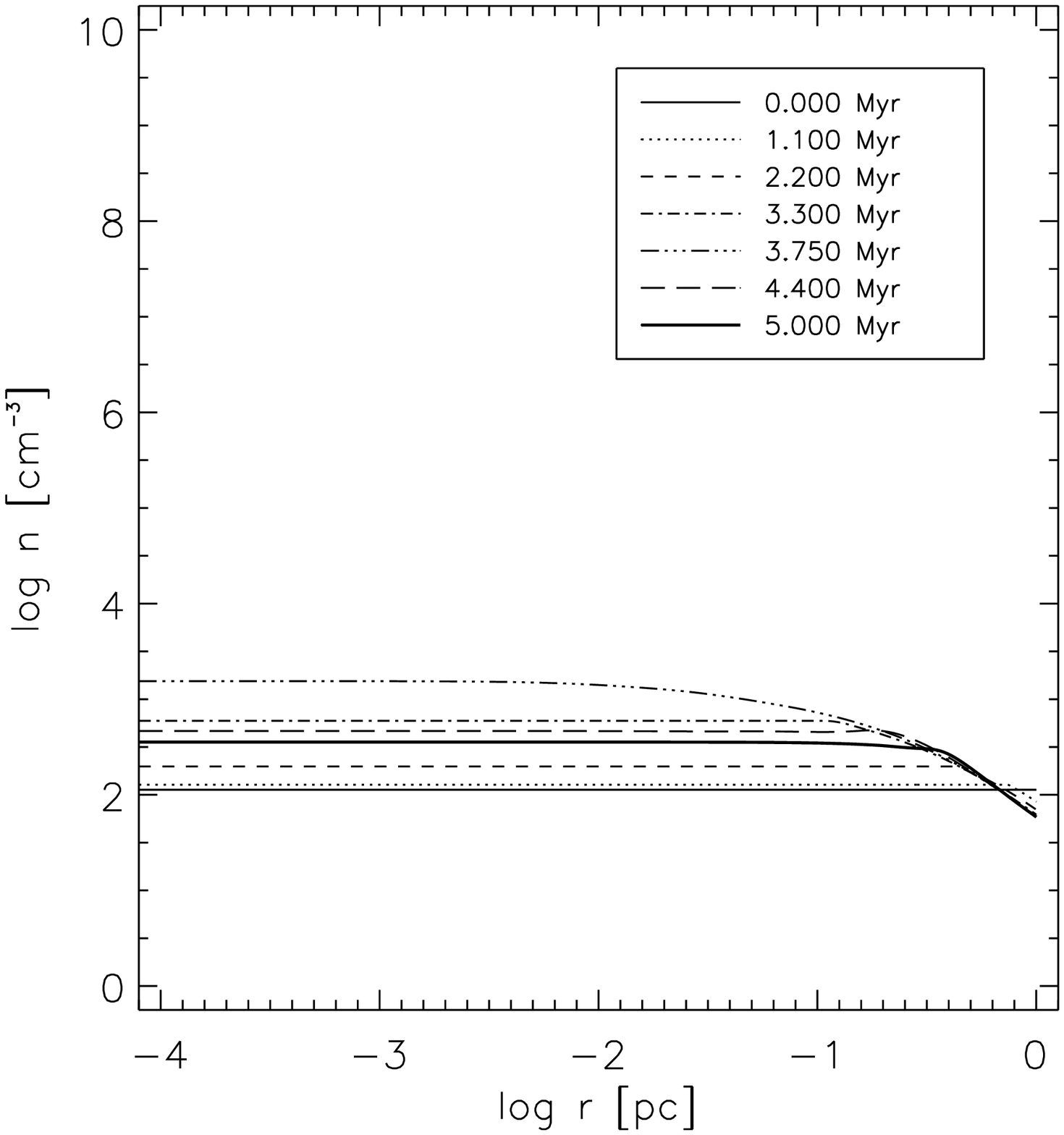}
\caption{
    Spontaneous collapse of the cloud without velocity impulse.
  {\it Left:} when started with a constant density $n = 175.70 \pcc$,
  the central region of the cloud undergoes gravitational collapse
  after a small bounce off the center.
  {\it Right:} when started with a constant density $n = 112.72 \pcc$,
  the cloud bounces off its center and expands until it disperses.
  \label{zero_vel:fig}
}
\end{figure*}

\section{THE SIMULATIONS}
\label{sec:results}

\subsection{Spontaneous collapse}
\label{sec:spont_col}

In order to study the effect of velocity fields in inducing
the collapse of molecular cloud cores, we first
need to determine when they can collapse under the influence of their
self-gravity alone. Because of the adopted 
spherical geometry (the usual Jeans analysis is applicable
to sinusoidal perturbations in plane parallel geometry),
the critical density $\rho_{\rm c}$ and mass (which we refer to as the
effective Jeans mass) at which the core collapses may differ slightly
from the standard Jeans values, and so we determine them here
numerically. We set $v_0 = 0$ and let the simulation run for $10
\Myr$ with a series of different initial densities.

As the simulations are started, self-gravity causes the cloud to
begin contracting, increasing its mean density
(see fig. \ref{zero_vel:fig}).
At some point, the pressure gradient in the inner parts stops this
process and the contraction is reversed (the cloud ``bounces''
momentarily). 
If the cloud's mass is large enough, self-gravity takes over again,
the expansion is also reversed and the cloud collapses;
otherwise, the
expansion continues until the simulation ends.
It is found that an initial density value of $160 \pcc$ yields a
collapsing core, while a $2\%$ lower density does not;
therefore, we take the critical density as
$\rho_{\rm c} = 160 \pcc$. At this density, the mass
in our numerical box (of radius $R=1$ pc) is $\Mbox (\rhoc) = 39.1 M_\odot$.
For comparison, the mean density for which the standard Jeans length equals
the diameter of the numerical domain ($2\pc$) is $\bar\rho = \rhoj \equiv \pi
\cs^2/G\LJ^2 = 125 \pcc$.

In the light of this result, we define the \emph{effective} Jeans mass as the
spherical Jeans mass (i.e., a sphere with diameter equal to the Jeans
length at mean density $\bar\rho$) times a fudge factor $A$ so that the
product equals the box's mass at the empirical critical density:

\begin{eqnarray}
  \MJeff &=& A\, \frac{4\pi\rhoc}{3} \left(\frac{\LJ}{2}\right)^3
                 \nonumber \\
      &=& A\, \frac{\pi^{5/2}}{6} \left(\frac{\cs^2}{G}\right)^{3/2}
          \rhoc^{-1/2},
\end{eqnarray}

\noindent
where $\kB$ is the Boltzmann's constant, $T$ is the temperature,
$\mu = 2.36~m_{\rm H}$ is mean particle mass,
and $G$ is the gravitational constant.
By setting $\MJeff = \Mbox$, we obtain $A = 1.45$. For comparison, the
standard Jeans mass at $\rhoc$ is $M_{\rm J} = 27.0 M_\odot$,
and the
BE mass \citep{Ebert55, Bonnor56} is $M_{\rm BE} = 1.18 \cs^3
/\left(G^3 \rhoc\right)^{1/2} = 10.0 M_\odot$.

%-----------------------------------------------------------------

\begin{figure*}[!t]
%\plotone{figures/J90-S2-r0.33.eps}
\plotone{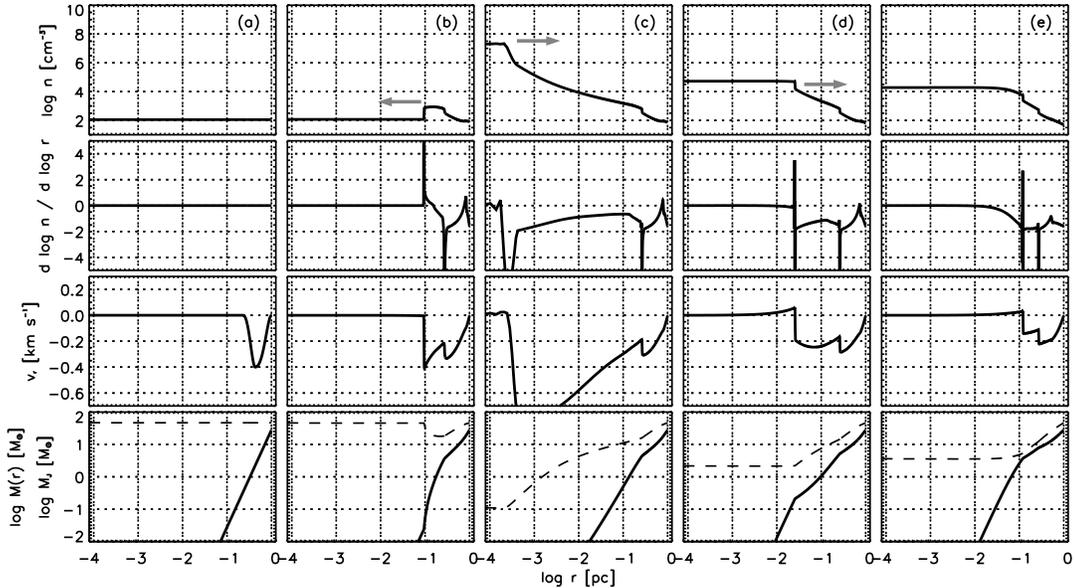}
\caption{
  Evolution of simulation S1.
  Each row shows the evolution (down from the top)
  of density ($n$),
  logarithmic density slope ($\dif \log n / \dif \log r$),
  velocity ($v_r$), mass internal to radius $r$ ($M(r)$, solid line)
  and effective Jeans mass ($\MJeff$, dashed line) at
  $0.000$ (column {\it a}), $0.625$ ({\it b}), $0.775$ ({\it c}),
  $0.925$ ({\it d}), and $1.500 \Myr$ ({\it e}).
  Arrows show the direction of motion of the shocks.
  [{\it See {\tt
  http://www.astrosmo.unam.mx/\~{ }g.gomez/publica/f2.mpg} for
  an mpeg animation of this figure.}]
  \label{S1:fig}
}
\end{figure*}

\begin{figure*}[!t]
%\plotone{figures/J90-S2-r0.67.eps}
\plotone{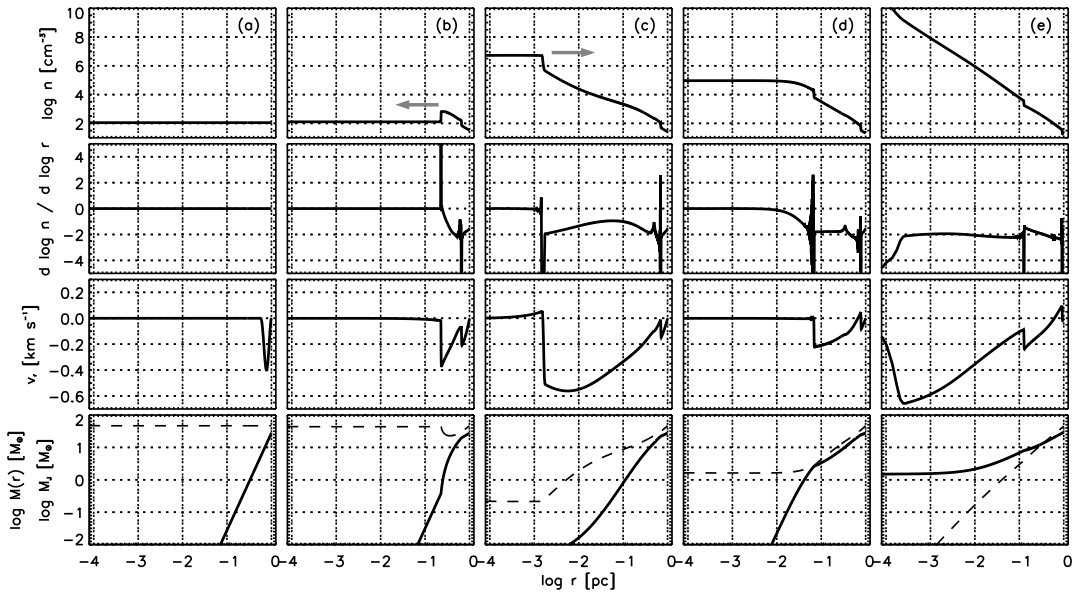}
\caption{
  Evolution of simulation S2 at
  $0.000, 1.125, 1.525, 2.000$ and $2.625 \Myr$.
  (See caption in fig. \ref{S1:fig}).
  [{\it See {\tt
  http://www.astrosmo.unam.mx/\~{ }g.gomez/publica/f3.mpg} for
  an mpeg animation of this figure.}]
}
\label{S2:fig}
\end{figure*}

\subsection{Cores formed by ram-pressure}
\label{sec:induced_col}

Although a large number of simulations were performed, our
discusion will focus on two of them, respectively representative of
non-collapsing and collapsing cases.
We shall call S1 the simulation with the initial velocity impulse
at $r_0 = 0.33 \pc$, while simulation S2 places the impulse at
$r_0 = 0.67 \pc$.
Both simulations have the same velocity amplitude
($v_0=0.4 \kms = 2 c_s$)
and sub-critical initial density
($112.7 \pcc \approx 0.7 \rho_{\rm c}$),
meaning that in the absence of compressive motions, both simulations
would simply expand away.

The evolution of simulation S1 is shown in Figure \ref{S1:fig}. This
figure respectively shows, as a function of radius, the density, the
logarithmic slope of the density radial profile, the velocity, and the
core's mass ({\it solid line}) and Jeans mass ({\it dashed line}) inside the
radius, in the four rows from top to bottom at selected 
times (left to right columns).
Shortly after the starting time ($t \approx 0.28$ Myr),
a shell bounded by two shocks appears on the inner side
of the initial 
velocity pulse, at $\log r \approx -0.7$ ($r \approx 0.2$ pc).
The formation of these two shocks and the shell between them is due
to the middle parts of the compressive wave, which have the highest
velocities, catching up with the frontal parts of the wave, causing
the formation of a shock, which splits into two shocks receding from
each other and leaving the shock-bounded layer in between.

The innermost one of these shocks propagates inward, 
leaving a large amount of
inflowing mass behind it  (fig. \ref{S1:fig}{\it b}). 
As it travels toward the center of the cloud ($t \approx 0.77$ Myr),
geometrical focusing dramatically increases the internal density,
lowering the effective Jeans mass of the inner parts of the cloud
(fig. \ref{S1:fig}{\it c}).  As the shock bounces off the center
and expands outwards, the shocked gas behind it is left at uniform
density and at essentially zero velocity; that is, a
\emph{quiescent core is formed}, 
with the shock-bounded shell dubbing as the
core's envelope
({\it d}).  The gas from the shock-bounded shell continues
to fall in, being incorporated into the
quiescent core as it passes through the inner shock. 
Although the mass of the quiescent core
increases, its density is somewhat
lowered because of a mild expasion
of the compressed region.  As a result, the  Jeans
mass becomes larger in the innermost parts of the core, and
decreases close to the shock-bounded layer and within it [compare
panels {\it (d)} and {\it (e)} { of fig.\ \ref{S1:fig}}].
As the core acquires more mass,
its density profile starts to deviate from being uniform, and
to approach that of a truncated BE sphere.
However,
in this simulation the mass of the inner core never becomes equal
to $\MJeff$ at any radius, and the uniform-density core begins
to expand indefinitely, developing positive velocities at the
outermost regions first (fig.\ 2{\it e}).

In simulation S2, the early evolution is quite similar to that of S1.
At $t\approx 1.5$ Myr, the inner shock
bounding the layer bounces off the center of the 
core (fig. \ref{S2:fig}{\it c}) and begins traveling outwards. But then,
some 0.5 Myr later ($t \approx 2.0$ Myr), the amount of mass in the
core finally becomes equal to $\MJeff$ ({\it
d}) at $r \approx 0.07$ pc, and from that moment on,
collapse ensues, culminating with the formation of a singularity
at $t \approx 2.6$ Myr ({\it e}). It is interesting that at $t \approx
2.0$ Myr, the mean density in the quiescent core is $n \approx 2.85
\times 10^4 \pcc$, implying a free-fall time $\tau_{\rm ff} \approx 0.2$
Myr, less than half the time the actual collapse takes.
There are several reasons possibly responsible for the discrepancy
with the observed collapse time of 0.6 Myr (from $t=2$ Myr to $t=2.6$
Myr).
For example, it was already
noted by \citet[][ Appendix C]{Larson69} that the actual collapse
time lasts nearly 1.5 times the free-fall time, because the pressure
gradient is never negligible. The remaining difference
is probably due to the imprecisions introduced by considering
the mean density rather than the detailed radial distribution, and
the fact
that the core is increasing its mass, so that the instability sets in at
an undetermined radius.

A very interesting feature of both
simulations is the fact that the 
density structure of the core+envelope system resembles
a BE sphere during the period over which the
shock front is traveling outwards
(figs. 
\ref{S1:fig}{\it e}  and \ref{S2:fig}{\it d}), at which times
the innermost parts of the core 
have a nearly constant density and the shock-bounded
layer approaches an $r^{-2}$ density profile.
After the formation of the singularity at the center, the $r^{-2}$
density profile extends throughout the core, similarly to an SIS profile
(fig. \ref{S2:fig}{\it e}).

It is worth remarking that, even though the 
idealized geometry and initial conditions
adopted in this paper should not have a strong impact
on the general qualitative behavior of
these simulations,
the above quoted sizes and time-scales
\emph{are} expected to depend on these details.
In fact, the outcome of the simulation (collapse or 
re-expansion) already 
depends sensitively on the parameters of our idealized simulations.
For example, the only difference between simulations S1
and S2 is the initial position of the compressive pulse. Moreover, a
simulation similar to S1 but started with $v_0 = 3 c_s$ and
a lower initial density ($88.8 \pcc$)
collapses $0.4 \Myr$ after the initial shock bounces off the center.
That is, a lower initial mean density can be counterbalanced by a larger
compressive Mach number, in the quest for inducing collapse.
Nevertheless, this simulation
still goes through 
the same qualitative evolution as simulation S2.
For these reasons,
the results presented in this section can be regarded as a
qualitative description of the formation and evolution of
molecular cloud cores,
while more quantitative analysis requires more realistic simulations
involving multi-dimensional compressions, and it is left for future
work.

%-----------------------------------------------------------------

\section{DISCUSSION}
\label{sec:discussion} 

\subsection{Generality of shock-bounded self-gravitating structures}
\label{sec:shck_bndd_struct} 

The formation of \emph{growing} shock-bounded structures is not
exclusive to the spherical symmetry used in this paper. Planar
compressions are generally known to produce shock-bounded layers in both
isothermal and radiatively-cooling flows. The one-dimensional
plane-parallel problem is equivalent to that of a shock front hitting a
wall and then reflecting
off it, and thus by construction the gas between the wall and the
shock is at rest, with the shock front receding from the wall at the
post-shock velocity of the flow. The shocked layer increases its mass as
gas from the incoming flow is incorporated into it after crossing the
bounding shock.
The shock bounded layer is thus the planar equivalent of our
spherical-shock-bounded quiescent core.
The plane-parallel problem has been worked out analytically in 1D and
numerically in two or three dimensions by \citet{FW06}
for the isothermal case
and by \citet{HP99} and \citet{VS_etal06} for the thermally bistable
case.
The main difference with the spherical case is that plane-parallel
compressions in thermally bistable media (such as
the warm HI gas) \emph{can} induce gravitational collapse
\citep{Hunt_etal86, VGJBG07} because in
this case the flow behaves effectively as if having $\gamef < 0$, while
in an isothermal medium focused (i.e., multidimensional)
compressions %($\nu > 2$)
are necessary because in this case $\gamef =1$
(see \S\ref{sec:focusing}).
However, the planar and the
spherically symmetric cases are qualitatively similar in that both
involve the formation of a shock-bounded structure that grows in mass by
accretion through the shock until it becomes gravitationally unstable
and begins contracting. In the planar compression case, this process has been
modeled numerically by various workers \citep[e.g.,][]{Hunt_etal86, VGJBG07}.

%-----------------------------------------------------------------

\subsection{Implications} \label{sec:implication}

The evolution of simulation S2 has a number of interesting important
implications, which we now discuss.

First, a compressive wave (or a negative-divergence velocity field)
does not directly induce the collapse of an initially sub-Jeans core.
The collapse happens only
if at some point in the evolution the mass becomes larger
than the Jeans mass.  In all the collapsing
simulations we have performed here, this occurs only
after the resulting shock front has rebounded
off the center, traveled
outwards, and incorporated a large enough amount of mass
into the central core, so that a ``traditional'' Jeans
criterion for collapse [$M(r) > \MJeff$]  is satisfied there. Since the
material behind the shock is left at zero velocity, \emph{no turbulent
support is ever at play there}. That is, \emph{the collapse does not occur
because turbulence is dissipated in the core,}  as it is often believed,
but rather because the growing core eventually reaches the effective
Jeans mass.
Moreover, as the shock
continues to move outwards, the size of the region acquiring the
effective Jeans mass increases, so that the determination of the mass
that is subsequently
incorporated into the collapse happens ``on the spot'' in a highly
fortuitous manner.

Second, a near $r^{-2}$ density profile is 
approached at late times in the infalling envelope 
around the central core, both in
collapsing and non-collapsing cases. 
The central 
core, in turn, evolves from a near-flat density profile to that of a
truncated BE sphere as its mass increases. The central core
and the envelope are separated by the outwards-traveling shock, which
is, however, very mild, with a Mach number very close to unity.
Thus, the
core+envelope system may easily be taken for a single structure with a
density profile resembling that of a BE sphere.

Third, the velocity profile of the central core at all
times after it has formed,
has nearly zero velocity 
throughout.  This provides a physical basis for the
existence of quiescent (subsonic non-thermal velocity dispersion) and
{\it coherent} (non-thermal velocity dispersion nearly constant
through the core) cores, which in the turbulent model of
core formation are the stagnation points of the
turbulent flow in molecular clouds \citep{kle05}.

Fourth, there is a time delay between the formation of the core
and its gravitational collapse.
The quiescent core grows from the center as the shock moves
outwards incorporating mass into the central shocked region.
This process cannot happen instantaneously, but rather requires a finite
time until the core's mass equals the effective Jeans mass. In our
simulations, roughly 1 Myr spans from the moment of central core
formation  to the development of a singularity at the center, with
roughly half of it being spent without any tendency to collapse.
This time delay naturally 
explains the high frequency of observed starless cores
with BE-like profiles.

Fifth, it is important to remark that even though our quiesecent cores
are morphologically similar to BE spheres, they are dynamically very
different: they are \emph{not}
confined by the thermal pressure of a hot, tenuous medium, but instead
are confined by the ram-pressure of
the inflowing gas from the envelope, and growing in size and mass
accordingly, until they become dominated by gravity, at which point they
engage into collapse.

Sixth, and finally, it appears that the whole evolution is not very
amenable to a similarity solution because: a) the initial velocity pulse
is finite, so that the external flow is not rescalable. b) The inner shock
bounding the shock-bounded shell hits the center and bounces back
towards the exterior so the time of collision at the center breaks
the self-similarity.
c) The central core gradually increases its
self-gravity and eventually may become gravitationally
unstable, a process that continously transforms the core's density
profile from
uniform to being BE-like, first stable and then
unstable. Similarity solutions may be most applicable
\emph{after} the formation of the central singularity, as originally
suggested by \citet{Shu77}.

%-----------------------------------------------------------------

\subsection{Comparison with previous work}
\label{sec:comparison} 

It is interesting to put the results of our numerical simulations
in context with those of previous studies.
The main difference is that our simulations
have investigated the \emph{formation} of the cores in addition to their
subsequent collapse, in order to study whether BE-like structures
can be spontaneously
produced out of supersonic turbulent compressions in isothermal
molecular clouds. Thus, in particular, our study sheds light on the
realizability of the initial conditions used by previous works.

Our results suggest that in fully isothermal molecular clouds (i.e.,
without a warm, tenuous interclump medium that can stabilize a density
enhancement), collapsing structures formed by 
random turbulent compressions in the medium morphologically resemble BE
spheres through a large fraction of their evolution, because they
consist of a central core and an infalling envelope
which, at late times after the formation of the
central core, has a density profile with a slope close to $-2$.
This extends previous results that plane-of-sky angular averaging and 
line-of-sight averaging cause the \emph{observed} density profiles to be
smoother that the actual ones and easily confused with BE ones
\citep{bp03, har04}.
Also at late times, near the
onset of gravitational collapse of the core, the latter also develops a
density profile close to that of a BE sphere, which connects with that
of the envelope. This means that, at the 
onset of gravitational collapse, our simulations favor
initial conditions for collapse such as those used by
\citet{FC93}, albeit with the 
added ingredient of a continuous accretion at the bounding shock.

The establishment of a near $r^{-2}$ density profile in
the envelope at late times is interesting in the context of the
discussion by \citet{Shu77}. He points out that the development of such
a profile requires that the initial motions in the outer regions of
collapsing cores be subsonic, so that all fluid parcels are in acoustic
contact with each other, and can therefore approach detailed mechanical
balance. In our simulations, however, this need for an initially
subsonic condition appears to be in contradiction with the supersonic
nature of the initial pulse. However, the acoustic
contact is restored in the envelope because it consists of shocked gas
that has been thermalized and thus initially subsonic. The fact that a
near-$r^{-2}$ profile develops in the envelope even in the
non-collapsing simulation can be understood because the compressive
pulse effectively removes the support for the outer layers, analogously
to the effect of an inside-out collapse.

Some authors have already studied spherically symmetric flows with
shocks in the context of protostar formation using
similarity methods \citep[e.g., ][]{SL04, LG06, LW06}. Similarity
studies are extremely useful in extracting the underlying asymptotic
behavior of real flows. Therefore, it is important to compare our
numerical solutions with existing similarity solutions of
self-gravitating clouds in the presence of shocks, in particular those
of \citet[][ herefater SL04]{SL04}, whose study most resembles our
numerical setup. These authors 
presented two possible classes of self-similar 
shocked flow in the context of the dynamical evolution of protostars,
depending on the 
asymptotic behavior of the solutions near the center of cloud. 
Their Class I solutions had negative (inflow) velocities ($\propto -
r^{1/2}$), a density 
profile $\rho \propto r^{-3/2}$, and finite mass as asymptotic limits at
$r \rightarrow 0$, while their Class II had positive (expansion)
velocity ($\propto r$), constant finite density and vanishing mass
($\propto r^3$) as the asymptotic behavior in the same limit. In both
classes, an outward-moving shock separates a collapsing (or
expanding) inner part and an accreting outer part. None of these
behaviors are seen in our simulations at any time. Their Class II is
similar to our solutions during the core-growth stage, in that it has a
uniform central density and an accreting outer part, which has a
counterpart in the infalling shock-bounded layer in our models. However,
 in our system the central core is neither
expanding nor contracting, but rather it is at rest. This difference is
most likely a consequence of self-gravity being neglegible in
our cores during the early stages of their evolution.
That is, unlike the SL04 solution, where
self-gravity is important at all radii and at all times,
in our simulations the relative importance of self-gravity increases
secularly with time, going from being zero at the time of core formation
to being dominant at the time when gravitational instability sets in.

Another recent study that is closely related to ours is that by
\citet{HWGA03}, who numerically investigated the effect of increasing 
the pressure external $\Pext$ to an initially stable BE sphere. These
authors found that slow rates of increase of $\Pext$ cause the sphere to
approach instability quasi-statically, but higher rates of increase
produced a compressive wave that triggers an outside-in collapse. It is
noteworthy, however, that they do not report the bounce of the
compressive wave from the center that we find. This is most probably 
because, in their case, the wave compresses a
previously-existing core that is in a (fragile) stable hydrostatic
equilibrium state, and so the role of the wave is to directly trigger
the collapse. Instead, in our case, the compressive wave 
\emph{forms} the core,
and adds mass to it until it becomes
gravitationally unstable and proceeds to collapse. Moreover, in the case
of \citet{HWGA03}, the mass of the core was fixed, being bounded by a
hot, tenuous medium, while in our case, the
fraction of the mass that is driven to collapse is determined ``in real
time'' by the interplay between the accreting gas and the outgoing shock
wave, and moreover the mass that becomes gravitationally unstable
increases with time, so the collapse proceeds ``inside out'' but over an
intermediate range of radii. Thus, we see that the choice of
equilibrium or out-of-equilibrium initial conditions and continuous or
discontinuous boundary conditions leads to very different patterns of
evolution. Which model applies best to actual turbulent molecular clouds
probably 
depends on whether they consist of a single, nearly isothermal molecular
phase (our model), or of a mixture of colder, denser molecular cloudlets
immersed in a more tenuous and warmer atomic medium
\citep{HI06}. Extensive theoretical and observational work, focusing
especially on the velocity structure of the cores, is needed
to decide on this issue. The recent results of \citet{Lee_etal07},
indicating the presence of a sharp infall velocity increase at $\sim
0.03$ pc from the centers of the starless cores L694-2 and L1197, would
seem to favor our dynamical scenario for the formation of the
cores. 

Finally, our results are fully consistent with the
scenario outlined by \citet{Whit_etal07}. These authors have foreseen the
formation of evolving BE spheres bounded and fed by the accretion of external
infalling material, which can collapse if the core eventually reaches
the BE mass. Although they restricted their discussion to the formation
of brown dwarf-producing cores, the simulations described here are seen
to be applicable to the formation of cores of arbitrary mass. It is
indeed likely that, as the core to be formed is of smaller mass, the
required compression  and focusing need to be stronger, as the initial
conditions will have sizes much smaller than the local Jeans length
(cf.\ \S \ref{sec:induced_col}).

%-----------------------------------------------------------------

\section{SUMMARY AND CONCLUSIONS} \label{sec:conclusions}

In this paper we have performed a numerical study of the formation of
dense cores by dynamical compressions in isothermal, non-magnetized
media, using simple one-dimensional calculations in spherical geometry.
Our results show that cores assembled by this process consist of a
central, quiescent core with density $10^5 \pcc$ that grows in mass and
size as it accretes mass 
from a surrounding envelope.
The quiescent core and the envelope are separated by a 
mild
shock with Mach number just above unity, and the accretion from the
envelope provides ram-pressure that confines the central quiescent,
growing core.
As the central core increases its mass, it passes first
through a neglegible-gravity, uniform-density stage, and later,
as self-gravity becomes important, it evolves into 
a ``pseudo BE-sphere'' stage.
If at some point in the evolution, the mass
of the core-envelope becomes lager than the Jeans mass, the core
proceeds to collapse. Otherwise, it begins to
reexpand.   
Even in collapsing cases, this process requires a
relatively long time to complete, taking $\sim 0.5$ Myr from the
first appearance of the central core to the time at
which it becomes
gravitationally unstable, and another $\sim 0.5$ Myr for the collapse to
produce a singularity at the center.

At all times after the formation of the central
core, the combined density structure of the
core+envelope system resembles that of a BE
sphere, since  it is flattened at the center,
and has an
envelope that approaches an $r^{-2}$ density profile at late
times. Moreover, the central core is quiescent at all times, except for the
very late stages of the non-collapsing case, in which the core begins to
expand.
Thus, the high observed frequency of BE-like profiles
\citep[e.g.][]{diFran07, LAL07}
{ and quiescent/coherent velocity dispersion structure
\citep{Myers83, Goodman_etal98, Cas02, Tafalla_etal02,
TMCW04, Schnee_etal07} is naturally accomodated in this scenario of dynamic
assembly of MC cores, as suggested also by studies of dense cores in
turbulent simulations of 3D, isothermal molecular clouds \cite{kle05}.
However,
the structures are not classical BE spheres, because they are confined
by ram-pressure, rather than by thermal pressure,
and are consequently accreting
mass and growing in mass, size, and self-gravitating energy, in a
process qualitatively similar to that described for the formation of
giant MCs by \citet{VGJBG07}.
In both cases, there is a secular evolution,
characterized by the mass increase of the cloud or core. 

The velocity structure of the cores  formed
in our simulations appears consistent with
recent radiative transfer models for the structure of cores L694-2 and
L1197 presented by \citet{Lee_etal07}, which exhibit a nearly zero
central velocity and a sharp rise at radii $\sim 0.03$ pc.
We plan to
carry out a radiative transfer study of the density and velocity
structures produced by our models in the near future, in 
order to perform detailed comparisons with observational studies based
on multi-tracer studies \citep[e.g.][]{LMP04} 
as well as on line-profile mapping of prestellar cores
\citep[e.g.,][]{Tafalla_etal98, TMCW04, LMT99, Lee_etal07, Schnee_etal07}.

%-----------------------------------------------------------------

\acknowledgements

We wish to thank S. Lizano and an anonymous referee
for useful comments.
This work has received financial support from
CRyA-UNAM to G.C.G., CONACYT grant U47366-F to E.V.-S, and
UNAM-PAPIIT grant 110606 to J.B.-P.
The research of M. S. was funded under the
Programme for Research in Third Level Institutions (PRTLI) administered
by the Irish Higher Education Authority under the National Development
Plan and with partial support from the European Regional Development
Fund.

%-----------------------------------------------------------------

{}


\begin{thebibliography}{}

\bibitem[Alves, Lada \& Lada (2001)]{ALL01}
Alves, J., Lada, C. \& Lada, E. 2001, Nature, 409, 159

\bibitem[Ballesteros-Paredes \& Hartmann(2007)]{BH07} 
Ballesteros-Paredes, J., \& Hartmann, L.\ 2007, Revista Mexicana de 
Astronomia y Astrofisica, 43, 123 

\bibitem[Ballesteros-Paredes et al.(1999)]{BHV99} 
Ballesteros-Paredes, J., Hartmann, L., \& V{\'a}zquez-Semadeni, E.\ 1999, 
\apj, 527, 285 

\bibitem[\BP\ et al.\ (1999)]{BVS99}
\BP, J., \VS, E. \& Scalo, J. 1999,
\apj, 515, 286

\bibitem[\BP\ et al.\ (2003)]{bp03}
\BP, J., Klessen, R. S., \& \VS, E. 2003
\apj, 592, 188

\bibitem[Bonnor (1956)]{Bonnor56}
Bonnor, W. B. 1956, MNRAS, 116, 351

\bibitem[Brice{\~n}o et al.(2001)]{Briceno_etal01} Brice{\~n}o, C., et 
al.\ 2001, Science, 291, 93 

\bibitem[Caselli et al.\ (2002)]{Cas02}
Caselli, Benson, Myers \& Tafalla 2002,
\apj, 572, 238

\bibitem[Chandrasekhar (1981)]{Chandra61}
Chandrasekhar, S.
1981, Hydrodynamic and Hydromagnetic Stability
(New York: Dover)

\bibitem[di Francesco et al.\ (2007)]{diFran07}
di Francesco, J., et al. 2007,
in Protostars and Planets V, ed. B. Reipurth, D.  Jewitt, \& K. Keil
(Tucson: Univ. Arizona Press), 17

\bibitem[Ebert (1955)]{Ebert55}
Ebert, R.\ 1955, Zeitschrift f{\"u}r Astrophysik, 36, 222 

\bibitem[Elmegreen (1991)]{Elm91}
Elmegreen, B. G. 1991, in The Physics of Star Formation and Early
Stellar Evolution, ed. C. J. Lada \& N. D. Kylafis
(Dordrecht: Kluwer), 35

\bibitem[Elmegreen (1993)]{Elm93}
Elmegreen, B. G. 1993 \apj, 419, L29

\bibitem[Elmegreen(2000)]{Elmegreen00} Elmegreen, B.~G.\ 2000, 
\apj, 530, 277 

\bibitem[Folini \& Walder (2006)]{FW06}
Folini, D. \& Walder, R. 2006, \aap, 459, 1

\bibitem[Foster \& Chevalier (1993)]{FC93}
Foster, P. N. \& Chevalier, R. A. 1993,
\apj, 416, 303

\bibitem[Galv\'an-Madrid et al.\ (2007)]{GVKB07} Galv\'an-Madrid, R.,
\VS, E., Kim, J., \BP, J. 2007, ApJ submitted (astro-ph/07043587)

\bibitem[Goodman et al.(1998)]{Goodman_etal98}
Goodman, A.~A.,
Barranco, J.~A., Wilner, D.~J., \& Heyer, M.~H.\ 1998, \apj, 504, 223

\bibitem[Hartmann(2004)]{har04}
Hartmann, L.  2004,
in IAU Symp. 221, Star Formation at High Angular Resolution,
ed. R. Jayawardhana, M. G. Burton \& T.L. Bourke,
(San Francisco: PASP) 201 

\bibitem[Hartmann et al.(2001)]{HBB01} Hartmann, L., 
Ballesteros-Paredes, J., \& Bergin, E.~A.\ 2001, \apj, 562, 852 

\bibitem[Heitsch et al.\ (2001)]{HMK01}
Heitsch F., Mac Low, M.-M. \& Klessen, R. F. 2001, \apj, 547, 280

\bibitem[Hennebelle \& P\'erault (1999)]{HP99}
Hennebelle, P. \& P\'erault, M. 1999, \aap, 351, 309

\bibitem[Hennebelle et al.\ (2003)]{HWGA03}
Hennebelle, P., Whitworth, A. P., Gladwin, P. P., \& Andr\'e, P. 2003,
MNRAS, 340, 870

\bibitem[Hennebelle \& Inutsuka (2006)]{HI06}
Hennebelle, P. \& Inutsuka, S. 2006,
\apj, 647, 404

\bibitem[Hunter (1977)]{Hunter77} Hunter, C. 1977, ApJ 218, 834

\bibitem[Hunter \& Fleck (1982)]{HF82} Hunter, J. H., Jr. \& Fleck,
R. C., Jr. 1982, ApJ 256, 505 

\bibitem[Hunter at al.\ (1986)]{Hunt_etal86}
Hunter, J. H., Jr., Sandford, M. T., II, Whitaker, R. W., \& Klein,
R. I. 1986, \apj, 305, 309

\bibitem[Klessen et al.\ (2000)]{KHM00}
Klessen, R. S., Heitsch, F. \& Mac Low, M.-M. 2000, \apj, 535, 887

\bibitem[Kirk et al.\ (2005)]{KirkJ05}
Kirk, J. M., Ward-Thompson, D. \& Andr\'e, P. 2005,
\mnras, 360, 1506

\bibitem[Klessen et al.\ (2005)]{kle05}
Klessen, R. S., Ballesteros-Paredes, J., V\'azquez-Semadeni, E.,
\& Dur\'an-Rojas, C. 2005
\apj, 620, 786

\bibitem[Lada et al.\ (2007)]{LAL07}
Lada, C. J., Alv\'es, J. F. \& Lombardi, M. 2007,
in Protostars and Planets V, ed. B. Reipurth, D.  Jewitt, \& K. Keil
(Tucson: Univ. Arizona Press), 3

\bibitem[Larson (1969)]{Larson69}
Larson, R. B. 1969,
\mnras, 145, 271

\bibitem[Lee \& Myers(1999)]{Lee_Myers99} Lee, C.~W., \& Myers, 
P.~C.\ 1999, \apjs, 123, 233 

\bibitem[Lee et al. (1999)]{LMT99}
Lee, C. W., Myers, P. C. \& Tafalla, M. 1999, \apj, 526, 788

\bibitem[Lee et al. (2004)]{LMP04}
Lee, C. W., Myers, P. C. \& Plume, R. 2004, ApJS, 153, 523

\bibitem[Lee et al. (2007)]{Lee_etal07}
Lee, S. H., Park, Y.-S., Sohn, J., Lee, C. W. \& Lee, H. M. 2007,
astro-ph/0702330

\bibitem[Li et al.\ (2004)]{LNMH04}
Li, P. S., Norman, M. L., Mac Low, M.-M., \& Heitsch, F. 2004, \apj,
605, 800

\bibitem[Lou \& Gao (2006)]{LG06} Lou Y. Q., Gao, Y. 2006, MNRAS, 373, 1610

\bibitem[Lou \& Wang (2006)]{LW06} Lou Y. Q., Wang W. G., 2006, MNRAS,
372, 885 

\bibitem[McKee et al.\ (1993)]{MZGH93}
McKee, C. F., Zweibel, E. G., Goodman, A. A., \& Heiles, C. 1993, in
Protostars and Planets III, ed. E. H. Levy \& J. I. Lunine
(Tucson: Univ. Arizona Press), 327

\bibitem[Myers(1983)]{Myers83}
Myers, P.~C.\ 1983, \apj, 270, 105 

\bibitem[Ostriker et al.\ (1999)]{OGS99} Ostriker, E. C., Gammie, C. F.,
\& Stone, J. M. 1999, ApJ, 513, 259

\bibitem[Ostriker et al.\ (2001)]{OSG01} Ostriker, E. C., Stone, J. M.,
\& Gammie, C. F. 2001, ApJ, 546, 980

\bibitem[Penston (1969a)]{Penston69a}
Penston, M. V. 1969, MNRAS, 144, 425

\bibitem[Penston (1969b)]{Penston69b} Penston, M. V. 1969, MNRAS, 145, 457

\bibitem[Pringle et al.(2001)]{Pringle_etal01} Pringle, J.~E., Allen, 
R.~J., \& Lubow, S.~H.\ 2001, \mnras, 327, 663 

\bibitem[Sasao (1973)]{Sasao73}
Sasao, T. 1973, \pasj, 25, 1

\bibitem[Schnee et al.(2007)]{Schnee_etal07} Schnee, S., Caselli, P., 
Goodman, A., Arce, H.~G., Ballesteros-Paredes, J., \& Kuchibhotla, K.\ 
2007, ArXiv e-prints, 706, arXiv:0706.4115 

\bibitem[Shen \& Lou (2004)]{SL04} Shen Y., Lou Y. Q., 2004, ApJ, 611, L117

\bibitem[Shu (1977)]{Shu77} Shu, F. 1977, \apj, 214, 488

\bibitem[Stone \& Norman(1992)]{sto92}
Stone, J. M. \& Norman, M. L.  1992,
\apjs, 80, 753 

\bibitem[Tafalla et al.\ (1998)]{Tafalla_etal98}
Tafalla, M., Mardones, D., Myers, P. C., Caselli, P., Bachiller, R.,
\& Benson, P. J. 1998,
\apj, 504, 900

\bibitem[Tafalla et al.(2002)]{Tafalla_etal02} Tafalla, M., Myers, 
P.~C., Caselli, P., Walmsley, C.~M., \& Comito, C.\ 2002, \apj, 569, 815 

\bibitem[Tafalla et al.\ (2004)]{TMCW04} Tafalla, M., Myers, P. C.,
Caselli, P., \& Walmsley, C. M. 2004, A\&A, 416, 191

\bibitem[Tilley \& Pudritz (2004)]{TP04} Tilley, D. A. \& Pudritz,
R. E. 2004, MNRAS, 353, 769

\bibitem[Tilley \& Pudritz (2005)]{TP05} Tilley, D. A. \& Pudritz,
R. E. 2005, submitted to MNRAS (astro-ph/0508562)

\bibitem[Tohline et al.\ (1987)]{TBC87}
Tohline, J. E., Bodenheimer, P. H. \& Christodoulou, D. M. 1987,
\apj, 322, 787

\bibitem[\VS\ et al.\ (1996)]{VPP96}
\VS, E., Passot, T. \& Pouquet, A. 1996, \apj, 473, 881

\bibitem[\VS\ et al.\ (2003)]{VBK03}
\VS, E., \BP, J., \& Klessen, R. S. 2003, \apj, 585, 131

\bibitem[\VS\ et al.\ (2005)]{VKSB05}
\VS, E., Kim, J., Shadmehri, M. \& \BP, J. 2005,
\apj, 618, 344

\bibitem[\VS\ et al.\ (2006)]{VS_etal06}
\VS, E., Ryu, D., Passot, T., Gonz\'alez, R. F. \& Gazol, A. 2006,
\apj, 643, 245

\bibitem[\VS\ et al.\ (2007)]{VGJBG07}
\VS, E., G\'omez, G. C., Jappsen, A. K., \BP, J.,
Gonz\'alez, R. F., \& Klessen, R. 2007,
\apj, 657, 870

\bibitem[Ward-Thompson et al.\ (2007)]{WardT07}
Ward-Thompson, D., Andr\'e, P., Crutcher, R., Johnstone, D., Onishi,
T., \& Wilson, C. 2007,
in Protostars and Planets V, ed. B. Reipurth, D. Jewitt, \&
K. Keil (Tucson: Univ. Arizona Press), 33

\bibitem[Whitworth et al.\ (2007)]{Whit_etal07}
Whitworth, A., Bate, M. R., Nordlund, \AA, Reipurth, B. \& Zinnecker
H. 2007,
in Protostars and Planets V, ed. B. Reipurth, D. Jewitt, \&
K. Keil (Tucson: Univ. Arizona Press), 459

\bibitem[Williams et al.\ (1999)]{WMWdF99} Williams, J. P., Myers,
P. C., Wilner, D. J., \& di Francesco, J. 1999, ApJ, 513, L61

\bibitem[Zuckerman \& Palmer (1974)]{ZP74}
Zuckerman, B. \& Palmer, P. 1974,
\araa, 12, 279

\end{thebibliography}
\end{document}